\documentclass[twocolumn,prl,aps,showpacs,amsmath,amssymb]{revtex4}

\usepackage[dvips]{graphicx}
\usepackage{dcolumn}
\usepackage{bm}

\newcommand{\xv}{{\mathbf x}}

\newcommand{\cv}{{\mathbf c}}

\newcommand{\Tr}{{\rm Tr}}

\newcommand{\xh}{{\hat{x}}}

\def\Lm{{{\mathbf \Lambda}}}
\def\xld{{r}}
\def\dcl{\xi_{\rm den}}
\def\pcl{\xi_{\rm perc}}
\def\tcl{\xi_{\rm ten}}

\def\matg{{\rm\bf g}}
\def\tilg{\tilde{\rm\bf g}}
\def\dgp{{g}}
\def\ten{S}
\def\denlabel{D}
\def\xlilabel{X}
\begin{document}
\preprint{APS/123-QED}
\title{Scaling of Entropic Shear Rigidity}
\author{Xiangjun Xing, Swagatam Mukhopadhyay, Paul M. Goldbart}
\affiliation{Department of Physics, 
University of Illinois at Urbana-Champaign, 
1110 West Green Street, Urbana, Illinois 61801-3080, U.S.A.}
\date{October 13, 2004} 

\begin{abstract}
%
%
The scaling of shear modulus near the gelation/vulcanization transition 
is explored heuristically and analytically.  It is found that in a dense 
melt the effective chains of the infinite cluster have sizes that scale 
{\it sub-linearly\/} with their contour length.  Consequently, each chain
contributes $k_{\rm B} T$ to the rigidity, which leads to a shear modulus 
exponent $d\nu$.  In contrast, in phantom elastic networks the scaling 
is {\it linear\/} in the contour length, yielding an exponent identical 
to that of the random resistor network conductivity, as predicted by 
de~Gennes. 
For non-dense systems, the exponent should cross over to $d\nu$ when the 
percolation correlation length 
is 
much larger than the 
density-fluctuation length. 
\end{abstract}




\maketitle

\noindent 
{\it Introduction\/}---Gelation and vulcanization are continuous 
phase transitions from liquids to random solids, caused by the 
introduction of chemical (i.e.~permanent) crosslinks; see, e.g., 
Ref.~\cite{deGennes_polymer}. 
It is by now well established that the {\it geometrical\/} aspects 
of these transitions are correctly described by percolation 
theory~\cite{deGennes_polymer,ref:Peng+Goldbart2000,
ref:Peng+Goldbart+McKane2001}. 
Qualitatively speaking, both transitions---gelation/vulcanization and 
percolation---are controlled by the emergence and structure of an 
infinite cluster at the critical point. 

The elastic and thermodynamic properties of gels near the critical point 
are not as well established.  Of these, the most controversial is the 
scaling behavior of the static shear modulus $\mu$, which is defined in the 
following way.  
Consider a spatially homogeneous, volume-preserving shear deformation 
$\Lm$~\cite{comment_shear}, with $\det \Lm=1$.  Under such a deformation, 
the increase in the free energy of a gel is, to leading order, given 
by~\cite{ref:Treloar}
\begin{equation}
\label{shear_modulus}
\delta F = V \mu \left( \Tr\,\matg - d \right), 
\qquad 
\matg \equiv \Lm^{\rm T}\cdot \Lm, 
\end{equation}
where $\matg$ is the metric tensor and $d$ is the spatial dimensionality. 
The shear modulus is expected, on general grounds, to obey a 
scaling law near the critical point, 
\begin{equation}
\mu = \mu_0 \, \Theta(-r) \, |\xld|^f,
\end{equation}
where the control parameter $\xld$ measures (minus the) deviation of 
the cross-link density from 
criticality 
(i.e.~$\xld < 0 $
in the solid phase), and $\Theta$ is the unit step function.

Values reported for the exponent $f$, either from experiments or 
computer simulations, are rather scattered, and seem to suggest four 
different universality classes~\cite{Coniglio_2002}.  
For systems with {\em entropic} elasticity, which is the focus of 
this Letter, values of $f$ usually fall into one of two classes. 
Firstly, most numerical simulations~\cite{simulations} involving 
phantom networks, as 
well as many gelation experiments, suggest that $f$ has the same value 
as the conductivity exponent $t$ ($\approx 1.9$ in three dimensions) 
of random resistor networks (RRN), supporting a conjecture of 
de~Gennes'~\cite{deGennes_polymer}.  
A second class of experiments, as well as some simulations, support 
the scaling result $f = d\nu$ ($\approx 2.6$ for three dimensions), 
where $\nu$ is the 
    percolation 
correlation-length exponent, as proposed in Ref.~\cite{Daoud_Conniglio}.  
    The argument for the latter exponent 
is in the spirit of the classical theory of rubber elasticity 
(see, e.g., Ref.~\cite{ref:Treloar}), in which the elastic modulus 
acquires a contribution of $k_{\rm B}T$ per effective chain. 
The purpose of this Letter to use heuristic and analytical methods to 
outline a resolution of the long-standing apparent contradiction 
between the two plausible arguments, mentioned above, as well as the 
inconsistency across experimental and simulational data.

\begin{figure}[t]
\begin{center}
\includegraphics[width=7cm]{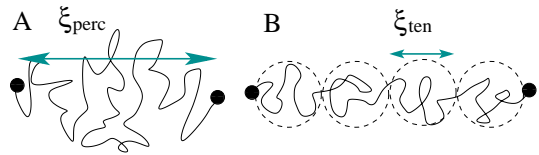}
\caption{The infinite cluster is a network of effective chains: 
(A)~In a dense melt each effective chain is one single thermal blob 
with typical size $\pcl$, and contributes $k_{\rm B}T$ to the 
shear rigidity. 
(B)~In a phantom network each chain is sequence of 
\lq\lq tension blobs\rq\rq\ (dashed circles), each of size $\tcl$. 
Each tension blob contributes $k_{\rm B}T$ to the shear rigidity.}
\label{cartoon}
\end{center}
\vspace{-5mm}
\end{figure}

\noindent{\it Heuristic reasoning\/}---Let 
us first consider gelation in a dense system with strong inter-particle 
repulsions, and let the system be near the critical point.  
Then the characteristic length-scale for density fluctuations $\dcl$ is 
much smaller than the percolation correlation length $\pcl$ (beyond which 
the infinite cluster is effectively homogeneous).  
Now let us invoke the ``nodes-links-blobs'' picture~\cite{NLB_picture,ref:urge}, 
in which the incipient infinite cluster is a network of effective chains, 
connected to one another at effective vertices, called nodes.
For our purposes, it is adequate to treat the effective chains as 
quasi-one-dimensional objects having some average thickness. 
%
The end-to-end displacement of these effective chains has a certain 
distribution with a characteristic length-scale, which is presumably
identical to the percolation correlation length $\pcl$. 
In the absence of external stress, we expect these chains to exhibit a type 
of random walk~\cite{comment_randomwalk}, owing to thermal fluctuations. 
Therefore, the end-to-end displacements should scale  
{\it sub-linearly\/} with the contour length. 
In the language of polymer physics, {\it every effective chain constitutes a 
single ``thermal blob\rq\rq\/}; see Fig.~\ref{cartoon} and Ref.~\cite{ref:RCtext}. 
Under a small shear deformation, each chain will be slightly stretched 
or compressed, thus contributing $k_{\rm B} T$ to the shear modulus.  
%
As there are roughly $\pcl^{-d}$ effective chains per unit volume, the 
shear modulus should scale as $k_{\rm B} T \pcl^{-d} \sim |\xld|^{d\nu}$, 
near the critical point.  
We note that the shear modulus has dimensions of energy per unit volume; 
hence, this scaling is also mandated by dimensional analysis, provided 
that $\pcl$ is the only important length-scale near the transition. 

Let us now consider typical numerical simulations of phantom networks, 
i.e., systems without inter-particle repulsions.  
The crucial observation, made above, that the equilibrium conformations of 
the effective chains are un-stretched random walks (i.e.~``thermal blobs'') 
with typical size $\pcl$ now breaks down, for the following reason.
In such simulations, the sol fraction (i.e.~the finite clusters) are 
usually removed completely, as they do not interact with the gel fraction.  
The resulting gel is very sparse, and would tend to collapse so as to 
maximize the entropy.  
To prevent this, the system size is usually fixed during shear deformation.  
Consequently, long effective chains are {\it strongly stretched\/}. 
The conformation of the infinite cluster is such that the net {\em entropic} 
force at each node vanishes, i.e., the cluster is {\it statistically\/} in 
mechanical equilibrium.  
Therefore, the mean tension $\ten$ carried by each chain has the same 
order of magnitude.  
As shown Fig.~\ref{cartoon}B, $\ten$ defines a length-scale, 
$\tcl \equiv k_{\rm B} T/\ten$, beyond which chain conformations are 
dominated by tension, and are thus effectively straight. 
By contrast, within $\tcl$ thermal fluctuations dominate, so that 
chain conformations are random walks.  
$\tcl$ is, by definition, the typical size of a ``tension 
blob''~\cite{tension_blob} for a polymer under tension $\ten$.  
Near the critical point, $\tcl$ is much smaller than $\pcl$, which 
diverges as $|r|^{-\nu}$.  
It follows that the number of tension blobs on each chain, as well as 
the chain's end-to-end displacement, scales {\it linearly\/} with its 
contour length in a phantom network. 
%
Under a shear deformation, every tension blob contributes $k_{\rm B}T$
to the overall shear rigidity~\cite{tension_blob}.  
Therefore, a typical chain, comprising many tension blobs, contributes 
a term to the total shear rigidity that is {\em proportional to its 
contour length}.
This should be contrasted with the case of melt, in which every chain 
contributes $k_B T$, {\em independent of its contour length}. 
In a coarse-grained description, we may replace every tension blob by a 
{\em mechanical} spring of natural length zero and force constant 
$k_{\rm B} T/\tcl^2$, 
without changing the elasticity.  
The resulting model is a randomly diluted network of {\em mechanical} 
springs of zero natural length {\em at zero temperature}, which can be mapped 
into the randomly diluted resistance network model~\cite{Tang_Thorpe}.  
In this mapping, the coordinates of nodes are mapped into voltages, and 
the shear modulus into the conductivity.  
Therefore the shear modulus of a randomly diluted entropic phantom network 
is equivalent to the conductivity of a random resistor network, 
as de~Gennes conjectured and many numerical simulations have supported. 


\noindent{\it Analytical reasoning\/}---We have extended the Landau theory 
for the elasticity of vulcanized matter\cite{ref:Peng+Goldbart2000} 
to the case of tunable repulsive interactions~\cite{vulcan_unpublished}.  
The relevant order parameter is the $(1+n)$-fold replicated particle 
density distribution, less its expectation value in the liquid phase:
\begin{eqnarray}
&&\Omega(\xh) \equiv
  \Omega(\xv^0,\xv^1,\ldots,\xv^n)
 =  -  
N\, V^{-(1+n)}
  \label{OP_def}\\ 
\!\!
&+&\sum\nolimits_{j=1}^N 
      \big[  \delta(\xv^0 - \cv_j)
\langle \delta(\xv^1 - \cv_j)\rangle
\cdots
\langle \delta(\xv^n - \cv_j)\rangle
\big]\!
    \nonumber            
\end{eqnarray}
Here, $\xh$ is short-hand for $(1+n)$ $d$-dimensional vectors
$(\xv^0,\xv^1,\ldots,\xv^n)$; 
$\cv_j$ labels the position of the $j^{\rm th}$ particle in the 
system, $\langle\cdot\rangle$ denotes a thermal average over the 
measurement ensemble~\cite{comment_ensembles}, and $[\cdot]$ denotes the 
average over the cross-linking realization (i.e.~the quenched disorder) 
as well as the thermal fluctuations of the preparation ensemble.  
%
$\Omega$ can be interpreted as giving the joint probability density 
function (p.d.f.) that a particle is located at $\xv^0$ in the preparation 
state and is later found at positions $\xv^1,\ldots, \xv^n$ in $n$ 
independent measurements in the measurement state. 
$\Omega(\xh)$ encodes a great deal of information about random solid state.  
In particular, the particle density fluctuations of the preparation and 
measurement states are respectively given by
\begin{subequations}
\begin{eqnarray}
&&                      \Omega^0(\xv)
                =               \big[{\small \sum}\delta(\xv - \cv_{j})\big]
                -               N/V,\\
&&                      \Omega^{\alpha}(\xv)
                =               \big[ \big\langle {\small \sum}
                                \delta(\xv-\cv_{j})\big\rangle \big]
                -               N/V,\\
&&\Omega^{\alpha}(\xv^{\alpha})
\equiv
\int\prod\nolimits_{\beta(\neq\alpha)=0}^n d\xv_{\beta}\,\Omega(\xh).
\end{eqnarray}
\end{subequations}
In the liquid phase, all particles are delocalized, i.e., 
$\langle\delta(\xv^{\alpha}-\cv_j)\rangle\equiv V^{-1}$ for all $j$.  
This ensures that the order parameter $\Omega(\xh)$ vanishes identically.  
In the gel phase, however, a certain fraction of particles belong to an 
infinite cluster and are localized. 
If such a particle were at position $\xv^0$ in the preparation state, 
it would 
fluctuate 
around the same point in the measurement state ({\it modulo\/} 
a global translation and rotation that are common to the system as a whole).  
This is captured by a nonzero expectation value of the 
joint p.d.f.~(\ref{OP_def}) in the gel phase.

The Landau free energy functional comprises one part concerning localization, 
and another describing density fluctuations. 
The localization part accounts for the effects of cross-links, and is given by
\begin{equation}
H_{\xlilabel}\!=\!\!
\!\int\! d\xh
\left\{
\!
\frac{\xld}{2}\,\Omega^2 
\!+\!\frac{K_0}{2} (\nabla^0 \Omega)^2 
\!+\!\frac{K}{2} {\small \sum_{\alpha=1}^n}
(\nabla^{\alpha} \Omega)^2  
\!-\!\frac{v}{3!} \Omega^3
\right\},
\label{H_X}
\end{equation}
where $\nabla^0$ and $\nabla^{\alpha}$ are, respectively, derivatives 
with respect to $\xv^0$ and $\xv^{\alpha}$.
The parameter $\xld$ controls the cross-link density and drives the 
transition to the random solid state, whereas $K_0$ and $K$ respectively 
measure the stretchability of polymer chains in the preparation and 
measurement states (the larger the $K$'s, the softer the chains).
The free energy cost for density fluctuations depends only on one-replica 
quantities, $\Omega^{\alpha}(\xv^{\alpha})$, and is 
\begin{equation}
H_{\denlabel}=
\frac{B_0}{2}\int d\xv^{0}\,\Omega^{0}(\xv^{0})^{2}+
\frac{B}{2} \sum_{\alpha=1}^n
\int d\xv^{\alpha}\,\Omega^{\alpha}(\xv^{\alpha})^2, 
\label{H_rho}
\end{equation}
where the non-negative parameters $B_0$ and $B$ are, respectively, the 
compressibility of the system in the preparation and measurement states. 
A large value of $B_0$ would ensure that the system is cross-linked in 
a state with almost uniform density profile, i.e., vanishingly small
fluctuations in $\Omega^{0}$.  
On the other hand, $B$ characterizes the repulsive interactions between 
particles in the measurement state. 
In a typical vulcanization experiment on a concentrated solution or melt, 
both $B_0$ and $B$ are large, so that the density remains essentially 
uniform across the transition. 
It is important, however, to realize that $B_0$ and $B$ are separately 
adjustable (as are $K_0$ and $K$), e.g., via tuning the 
solvent quality
before and after crosslinking. 
%
For 
gelation, the network formation process commonly lasts 
for an extended period.  
As a result, $B$ may differ from $B_0$, even by a large factor (e.g.~due 
to correlations built up during the course of the chemical reaction), 
even if all external physical conditions remain unchanged.  

In strong contrast, in typical numerical simulations of phantom 
systems, all polymers (or particles) are ascribed to lattice sites, and 
then cross-links are randomly introduced, connecting some neighboring 
particles.  
After removing the sol part, the system is allowed to relax at nonzero 
temperature, with intra-cluster repulsion completely ignored. 
This corresponds to a large positive value for $B_0$ but a {\it vanishing\/} 
value of $B$.  
As we shall soon see, it is this {\em qualitative} difference between $B_0$ 
and $B$ that is responsible for conductivity-like scaling of the shear 
modulus in phantom networks. 

\begin{figure}
\begin{center}
\includegraphics[width=8cm]{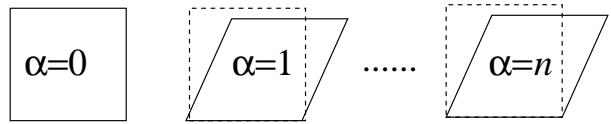}
\caption{ 
Shear deformation (full vs.~dotted lines) 
is applied to measurement but not preparation replicas.
After the coordinate transformation, the original boundaries 
are recovered at the cost of introducing a nontrivial metric tensor.}
\label{shear}
\end{center}
\vspace{-5mm}
\end{figure}

To study the elastic properties, we consider deforming the system 
{\it after\/} cross-links have been introduced.  
As shown in Fig.~\ref{shear}, this amounts to making an affine 
change $\Lm$ of the boundaries for measurement replicas 
($1$ through $n$), leaving the preparation replica intact.  
It is convenient to then make the linear coordinate 
transformation
\begin{equation}
\cv^\alpha \rightarrow \Lm \cdot \cv^{\alpha},
\hspace{3mm}
\xv^\alpha \rightarrow \Lm \cdot \xv^{\alpha},\,\,\,\,
\mbox{for $1 \leq \alpha \leq n$}, 
\end{equation}
which restores the original boundary conditions for the $n$ 
measurement replicas.  
Under this transformation, $H_{\denlabel}$ of Eq.~(\ref{H_rho}), 
is unchanged, provided we redefine $B$ appropriately.   
As for $H_{\xlilabel}$, Eq.~(\ref{H_X}), we find that all terms 
are invariant under the transformation (in the $n\rightarrow 0$ limit), 
except for the term with coefficient $K$, which becomes
\begin{equation}
\frac{K}{2}\sum_{\alpha=1}^{n}
\left(\nabla^{\alpha}\Omega\right)^2   
\longrightarrow
\frac{K}{2} \sum_{a,b=1}^d g^{-1}_{ab}
\sum_{\alpha=1}^n 
\nabla^{\alpha}_a \Omega\,
\nabla^{\alpha}_b \Omega\,,  
\end{equation}
where $\matg$ is the metric tensor defined in Eq.~(\ref{shear_modulus}).  

The effects of fluctuations, both thermal and quenched, are studied 
by averaging over all order-parameter configurations.  
Then the disordered-averaged physical free energy $[F]$ is related to 
the $(1+n)$-fold replicated partition function $Z_{1+n}$ by
\[
[F]=
-\lim_{n\rightarrow 0}
\partial_{n}
\ln Z_{1+n}= 
-\lim_{n\rightarrow 0}
\partial_{n}
\ln\!\int D\Omega \, e^{-H_{1+n}}, 
\]
where $H_{1+n}$ 
($\equiv  H_{\denlabel}+H_{\xlilabel}$) is the full 
effective\cite{comment_temperature} Hamiltonian; 
see Ref.~\cite{vulcan_Goldbart}.  

A renormalization-group (RG) analysis of the full model, described 
by $H_{1+n}$, will be presented in a future 
publication~\cite{vulcan_unpublished}.  
Here, we shall only present results for two limiting cases: 
(a)~$B_0 = B = +\infty$, 
i.e., vulcanization in an incompressible polymer melt; and 
(b)~$B_0 = +\infty$ but $B=0$, i.e., phantom networks.
We apply a momentum-shell RG transformation, thus integrating out 
short length-scale fluctuations recursively, whilst rescaling the 
order-parameter field $\Omega$ and spatial coordinates such that 
$K_0$ and $K$ remain unity.  
The renormalizations of $r$ and $v$ turn out to be independent of 
$B$ and $\matg$, i.e., the parameters describing the measurement 
ensemble.  To one-loop order we find: 
\begin{equation}
\frac{d\xld}{dl} = \xld\,
\left(2 + \frac{1}{6}v^2 \right) + \frac{v^2}{2},
\qquad
\frac{dv}{dl} = \frac{1}{2} \epsilon\,v - \frac{7}{4}v^3, 
\end{equation}
where $\epsilon \equiv 6-d$. 
For $d < 6$ there is a nontrivial fixed point at 
$(\xld^*,v^{*2}) = (\epsilon/14,2\epsilon/7)$. 
Correspondingly, the critical exponents $(\eta,\nu)$ are given by 
$(-\frac{1}{21}\epsilon,\frac{1}{2}+\frac{5}{84}\epsilon)$, 
which agree with results from the the $\epsilon$ expansion 
for the percolation transition~\cite{Lubensky_RRN}.

The flow of $\matg$ determines the elastic properties.  
Expressing $\matg$ as $\dgp\tilg$, where $\tilg$ has unit determinant 
[so that $\dgp^{d}=\det(\matg)$] we find that, regardless of the value 
of $B$, $\tilg$ does not flow.  
Therefore, renormalization of the metric $\matg$ is completely controlled 
by its determinant.  
For $B=B_0=+\infty$, we find the flow equation 
\begin{equation}
d\, \dgp / d \, l = 
(2/3) v^2
\left(1-\dgp^{3}\right)\dgp. 
\end{equation}
For $d<6$, $v^2\rightarrow v^{*2}=2\epsilon/7>0$, hence $\dgp$ flows to unity. 
Qualitatively, this implies a symmetry between the preparation and  
and measurement ensembles.  It also suggests that 
the correlation length $\pcl$ is the only relevant length-scale, 
in agreement with the preceding heuristic argument, that, in a dense melt, 
each effective chain constitutes a single thermal blob of typical size $\pcl$.  
Near the fixed point, the singular part of the free energy (i.e.~the elastic
free energy) has the scaling form
\begin{equation}
f_{\rm s}=
|\xld|^{d\nu}\,\psi_1(\matg^*)=|\xld|^{d\nu} 
\psi_1(\tilg).
\end{equation}
As the shear modulus is given by an appropriate derivative of $f_{\rm s}$ with 
respect to $\matg$ [cf.~Eq.~(\ref{shear_modulus})], we immediately see that it 
scales as $|\xld|^{d\nu}$. 

For the second case, viz., $(B_0,B)=(+\infty,0)$ we find
\begin{equation}
d\, \dgp/d\,l = (2/3) v^2 \dgp
\longrightarrow
 4 \epsilon \,g/21
\label{flow_g2}
\end{equation}
Now $\dgp$, and also the metric tensor $\matg$, are {\it relevant\/} 
near the percolation fixed point, with a positive crossover exponent $\phi_g$ 
of $4\epsilon/21$, echoing our heuristic argument that effective chains are 
strongly stretched in a phantom network.  
In general, the singular part of free energy should then have the scaling form 
\begin{equation}
f_{\rm s}=|\xld|^{d\nu}\,\psi_2(\matg/|\xld|^{\phi_g}). 
\label{f_s_2}
\end{equation}
For a pure shear with $\det(\matg)\equiv 1$, this $f_s$ must agree 
with Eq.~(\ref{shear_modulus}), up to a constant independent of $\matg$. 
Therefore the shear-modulus exponent is given by 
$d\nu-\phi_g = 3-\frac{5}{21}\epsilon$, 
which, to the same order in $\epsilon$, is identical to the conductivity 
exponent of a random resistor network~\cite{Lubensky_RRN}. 
 
This equivalence between critical exponents of phantom elastic networks 
and random resistor networks should in fact hold to all orders in 
$\epsilon$.  
To see this analytically, we note that setting $(B_0, B)=(\infty, 0)$ in 
$H_{\denlabel}$ is equivalent to setting $B_0 = B = 0$, {\it together with 
the hard constraint\/} $\Omega^{0}(\xv)\equiv 0$, which explicitly excludes 
configurations having nonzero density fluctuations in the preparation ensemble.  
The resulting model then becomes formally identical to the Harris-Lubensky 
formulation of the random resistor network problem~\cite{Lubensky_RRN}, 
with the $nd$ ($\to 0$) coordinates associated with the measurement ensembles 
$(\xv^1,\ldots,\xv^n)$ mapped onto the $D$ ($\to 0$) replicated voltages 
$\vartheta$, provided the stated limits are taken.
Thus, the pair of  systems are governed by identical RG equations and, 
hence, critical exponents.

Having established the two limiting cases, $B=\infty$ and $B=0$, it is 
natural to ask which one is the more stable.  
As $B$, like $\xld$, has na{\"\i}ve dimension $2$, it is always relevant 
near 6 dimensions.  
Therefore, if we keep $B_0$ large and tune $B$ to be small, the shear 
modulus exponent should cross over---from the conductivity one, $t$, to the 
incompressible system one, $d\nu$---when $|\xld|$ becomes smaller than $B$.  
This can be readily realized in a numerical simulation, if one were to 
retain the sol part of the system and turn on a small repulsion.  
In principle, this cross-over might also be observed in gelation experiments 
on non-dense solutions, provided $|\xld|$ is sufficiently small.

\noindent
{\it  Acknowledgments\/}---We thank 
E.~Fradkin and O.~Stenull for helpful 
discussions.  We acknowledge
support from 
NSF DMR02-05858 and 
DOE DEFG02-91ER45439.



\begin{thebibliography}{10}

\bibitem{deGennes_polymer}
P. G. de~Gennes, 
{\it Scaling Concepts in Polymer Physics\/}  
(Cornell University Press, Ithaca, NY, 1979).

\bibitem{ref:Peng+Goldbart2000}
W. Peng, {\em et al}, 
Phys. Rev. B {\bf 57\/}, 839 (1998); 
Phys. Rev. E {\bf 61\/}, 3339 (2000). 


\bibitem{ref:Peng+Goldbart+McKane2001}
H.-K. Janssen, O. Stenull,
Phys. Rev. E {\bf 64\/}, 026119 (2001);
W. Peng, P. M. Goldbart, A. J. McKane,
Phys. Rev. E {\bf 64\/}, 031105 (2001).

\bibitem{comment_shear}
A homogeneous deformation is described by a constant matrix $\Lm$, 
such that $\xv' = \Lm \cdot \xv$, where $\xv$ and $\xv'$ are 
the positions of mass points before and after 
deformation. 

\bibitem{ref:Treloar}
L. R. G. Treloar, 
{\sl The Physics of Rubber Elasticity\/} 
(Clarendon Press, Oxford, 1975).

\bibitem{Coniglio_2002}
For a brief overview, see E. Del Gado {\it et al\/}., 
\pre {\bf 65}, 041803 (2002) and reference therein.

\bibitem{simulations}
Recent numerical studies include: 
M.~Plischke {\it et al.}, \prl\ {\bf 80}, 4907 (1998),
\pre\ {\bf 60}, 3129 (1999);  O.~Farago and Y.~Kantor,
\prl\ {\bf 85}, 2533 (2000), \pre\ {\bf 62}, 6094 (2000). 

\bibitem{Daoud_Conniglio}
M. Daoud, A. Coniglio, 
J. Phys. A {\bf 14}, L-30 (1981).


\bibitem{NLB_picture} 
See, e.g., 
T. Nakayama {\it et al}., 
\rmp {\bf 66}, 381 (1994), 
and references therein.

\bibitem{ref:urge} 
Our arguments are not dependent on the details of the 
``nodes-links-blobs'' picture.  
%
In particular, the internal structure of the effective chains will 
not feature.

\bibitem{comment_randomwalk} 
Owing to the complicated internal structure of these chains and 
repulsive interactions, this random walk is not expected to be ideal.

\bibitem{tension_blob}
For a discussion of thermal blobs, tension blobs and force-extension 
curves for polymers, see, e.g., Ref.~\cite{ref:RCtext}.

\bibitem{ref:RCtext}
M. Rubinstein, R. H. Colby, 
{\sl Polymer Physics\/}  
(Oxford University Press, 2003).

\bibitem{Tang_Thorpe}
W. Tang, M. F. Thorpe, 
\prb {\bf 36}, 3798 (1987).

\bibitem{vulcan_unpublished} 
S.~Mukhopadhyay, X.~Xing, P.~M.~Goldbart, 
manuscript in preparation

\bibitem{comment_ensembles} 
The properties of gels and rubbers depend on both the state in 
which properties are measured and the state in which the system is prepared.  
%
A statistical-mechanical study of these materials necessarily involves two 
ensembles: a preparation one and a measurement one. 
%
The effect of cross-links is to couple these two ensembles.  
%
The quenched nature of the cross-links requires that the measurement ensemble 
be replicated $n$-times (with the $n\rightarrow 0$ limit ultimately taken).  
%
Early discussions of these issues can be found in 
Refs.~\cite{deGennes_polymer} and \cite{Deam_Edwards}.
%
A full account will be presented in a forthcoming 
manuscript~\cite{vulcan_unpublished}.

\bibitem{Deam_Edwards} 
R. T. Deam, S. F. Edwards, 
Phil. Trans. R. Soc. {\bf A, 280}, 317 (1976).

\bibitem{comment_temperature}
Recall that $H_{1+n}$ is an {\em effective} Hamiltonian obtained 
by integrating out the polymer degree of freedom of the 
original vulcanization model.  As a result, the
parameters in $H_{1+n}$ depend on the temperatures $T_0$ (before 
crosslinking) and $T$ (after crosslinking) in different ways.
As temperature is not the driving factor of the 
transition, we do not show it explicitly in the formalism.  
The only essential $T$ dependence comes from a factor of $T$
in the shear modulus, which is easily handled. 

\bibitem{vulcan_Goldbart} 
P.~M.~Goldbart {\it et al.}, 
Adv. Phys. {\bf 45}, 393 (1996).


\bibitem{Lubensky_RRN}
A. B. Harris, T. C. Lubensky, 
\prl {\bf 35}, 327(1975);
\prb {\bf 35}, 6964 (1987).



\end{thebibliography}

\end{document}